\documentclass[%
twocolumn,
superscriptaddress,
 amsmath,amssymb,
 aps,
prl,
]{revtex4-2}

\usepackage{graphicx}% Include figure files
\usepackage{dcolumn}% Align table columns on decimal point
\usepackage{bm}% bold math
\usepackage{color}

\begin{document}

\preprint{APS/123-QED}

\title{Broadband Magnomechanics Enabled by Magnon-Spoof Plasmon Hybridization}

\author{Yu Jiang}
\affiliation{ 
    Department of Electrical and Computer Engineering, Northeastern University, Boston, MA 02115, USA
}

\author{Jing Xu}
\affiliation{ 
    Department of Physics, University of Central Florida, Orlando, FL 32816, USA
}

\author{Zixin Yan}
\affiliation{ 
    Department of Electrical and Computer Engineering, Northeastern University, Boston, MA 02115, USA
}

\author{Amin Pishehvar}
\affiliation{ 
    Department of Electrical and Computer Engineering, Northeastern University, Boston, MA 02115, USA
}

\author{Xufeng Zhang}
\email{xu.zhang@northeastern.edu}
\affiliation{ 
    Department of Electrical and Computer Engineering, Northeastern University, Boston, MA 02115, USA
}
\affiliation{ 
    Department of Physics, Northeastern University, Boston, MA 02115, USA
}

\date{\today}

\begin{abstract}
Cavity magnomechanics is one important hybrid magnonic platform that focuses on the coherent interaction between magnons and phonons. The resulting magnon polarons inherit the intrinsic properties of both magnons and phonons, combining their individual advantages and enabling new physics and functionalities. But in previous demonstrations the magnon-phonon coupling either have limited bandwidth or cannot be efficiently accessed. In this work, we show that by utilizing the slow-wave hybrid magnonics configuration based on spoof surface plasmon polaritons (SSPPs), the coherent magnon-phonon interaction can be efficiently observed over a frequency range larger than 7 GHz on a YIG thin film device. In particular, the capability of the SSPPs in exciting short-wavelength magnons reveals a novel size effect when the magnons are coupled with high-over tone bulk acoustic resonances. Our work paves the way towards novel magnomechanical devices. 
\end{abstract}


\maketitle

%\tableofcontents

Hybrid magnonics has attracted significant attention in recent years due to its unique properties and application potentials \cite{Rameshti_PhysRep_2022,Harder_SSC_2018,Lachance_APE_2019,Bhoi_SSP_2020,YiLi_JAP_2020,Awschalom_IEEETransQuantEng_2021,Zhang2023Sep_MTE}. In hybrid magnonics, magnons -- quasi-particles representing the elementary collective excitation of spins -- strongly interact with other information carriers such as photons and phonons \cite{Huebl_PRL_2013,XufengZhang_PRL_2014,Tabuchi_PRL_2014,Goryachev_PRAppl_2014,LihuiBai_PRL_2015, JustinHou_PRL_2019,YiLi_PRL_2022,2024_PRL_Xu_slowwave,XufengZhang_SciAdv_2016, Potts_PRX_2021,2021_PRAppl_Xu_pulseEcho,An_PRB_2020,Hwang_PRL_2024,Shen_PRL_2022,Li2021Jun_APLM,Hatanaka_PRAppl_2022,An_PRX_2022}. These interactions give rise to magnon polaritons (through magnon-photon interactions) and magnon polarons (via magnon-phonon interactions), which inherit the intrinsic properties of both magnons and photons (or phonons). Such combination enables a plethora of intriguing phenomena, such as unidirectional invisibility \cite{YipuWang_PRL_2019,XufengZhang_PRAppl_2020} and hybrid magnonic exceptional points \cite{Zhang2017Nov_NC} and surfaces \cite{Zhang_PRL_2019}, enabling new opportunities for applications such as quantum transduction \cite{Tabuchi_science_2015Jul, Lachance-Quirion_Science_2020Jan, Lachance-Quirion2017Jul_SciAdv, Bolski_PRL_2020, Xu_PRL_2023May, Hisatomi_PRB_2016,NaZhu_Optica_2020} and dark matter detection \cite{Crescini_PRL_2020,Flower_PhysDarkUniv_2019}, all of which are unattainable in isolated systems. Notably, the coherent magnon-phonon interaction, known as magnomechanics\cite{XufengZhang_SciAdv_2016}, has demonstrated significant potential for both fundamental research and practical applications, such as ground state cooling \cite{Kani_PRL_2022,Ding2020Mar_JOSAB}, magnon-phonon entanglement \cite{Agarwal_PRL_2018_entanglement,Chakraborty_PRA_2023,Ding2022Oct_JOSAB}, and quantum-limited thermometry \cite{Potts_PRAppl_2020}.

Despite the wide band tunability of magnons and large frequency availability of phonons, the study of coherent magnon-phonon interaction has been restricted to small bandwidths in previous magnomechanical devices \cite{XufengZhang_SciAdv_2016,Potts_PRX_2021,2021_PRAppl_Xu_pulseEcho,An_PRB_2020}, which usually utilize narrow-band microwave resonators as transducers to access the interacting magnons and phonons. Although the bandwidth of magnon-phonon interaction can be expanded by using broadband transducers such as microstrips or coplanar waveguides, their efficiencies are highly limited, giving very low magnon extinction ratios, which can be less than 1 dB for thin-film YIG devices flipped on the transducer \cite{2024_PRL_Xu_slowwave}. This presents a significant challenge for the broader investigation of magnomechanics and its applications, particularly in the development of integrated magnonic devices \cite{Wang2020Dec_NE,Wang_PRAppl_2024}.

In this work, we show that by utilizing the enhanced magnon-microwave photon coupling on spoof surface plasmon polariton (SSPP) waveguides \cite{2024_PRL_Xu_slowwave}, the coherent interaction between magnons and high overtone bulk acoustic resonance (HBAR) phonons can be clearly observed over a wide frequency range, facilitating the broadband study of cavity magnomechanics. In particular, we noticed that the lateral confinement of the YIG magnonic resonators, which has been ignored in most previous studies, plays a pivotal role in determining the magnon-phonon coupling strength and has to be taken into consideration for short-wavelength magnon modes.

\begin{figure}[tb]
\includegraphics[width=\linewidth]{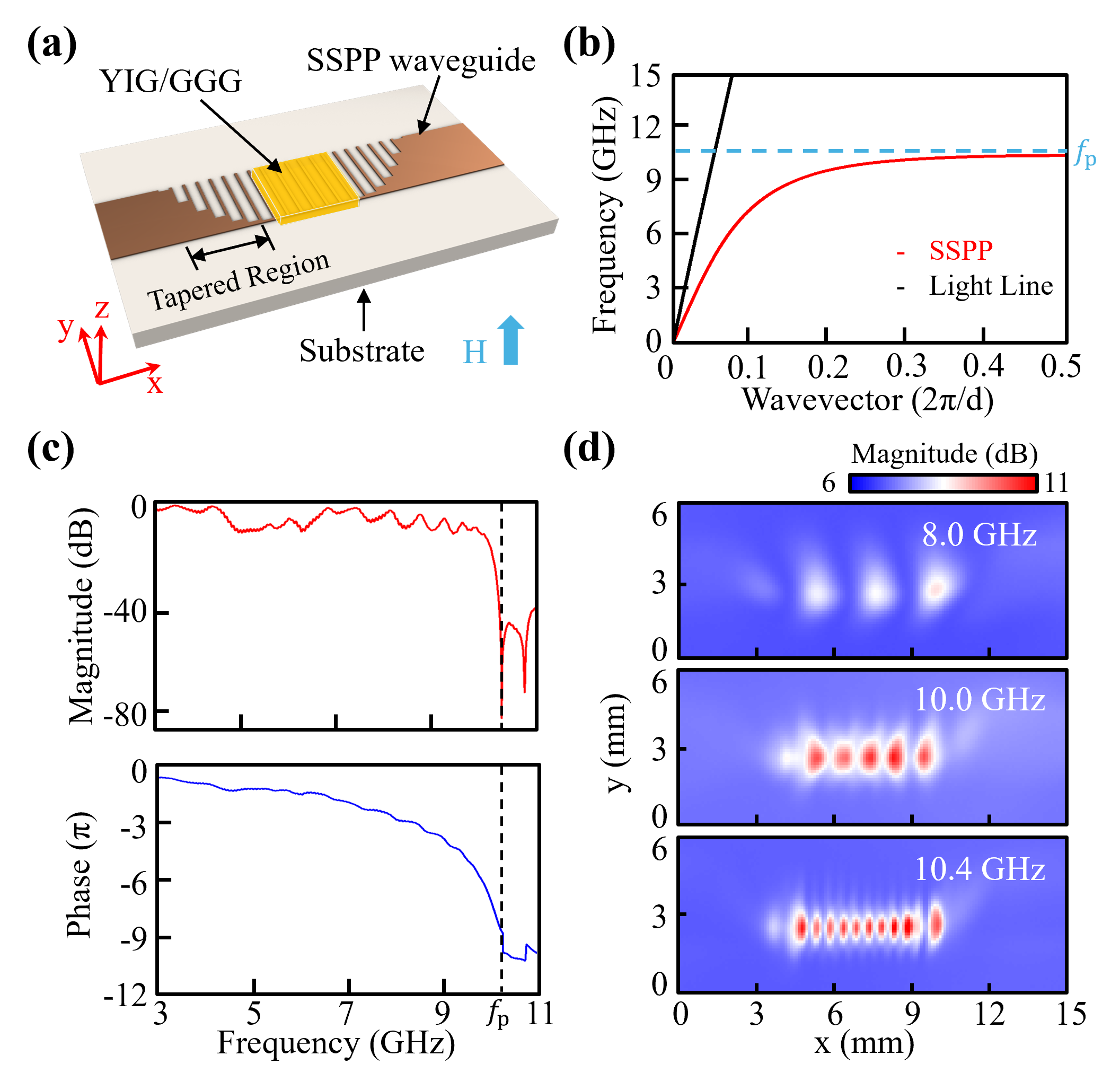}
\caption{(a) Schematic illustration of the broadband magnon-phonon device platform, with a YIG/GGG chip flipped on a SSPP waveguide. (b) Simulated dispersion curve of the SSPP waveguide. Light line is for the fundamental microstrip mode of the uncorrugated microstrip. (c) Measured transmission magnitude and phase of the fabricated SSPP waveguide. (d) Measured $h_z$ field distribution of the SSPP mode.}
\label{fig1}
\end{figure}

The device we used to study the broadband magnomechanics in schematically plotted in Fig.\,\ref{fig1}(a). The magnonic component supporting interacting magnons and phonons is a chip with a 200-nm thin film of yttrium iron garnet (YIG) epitaxially grown on a 500-$\mu$m gadolinium gallium garnet (GGG) substrate. Two different chips are used in our experiment, one $2.5\times 2.5$ mm$^2$ (Device 1)  and the other $3.0\times 3.0$ mm$^2$ (Device 2). A magnetic field is applied along the out-of-plane ($z$) direction to bias the YIG resonator, allowing the excitation of forward volume magnetostatic waves (FVMSWs) in the YIG thin film. The magnonic chip is flipped, with the YIG film facing down, on a SSPP waveguide, which has the form of a periodically corrugated microstrip. The SSPP waveguide is fabricated on a printed circuit board (Rogers TMM10i), which has a dielectric constant of 9.8 and a thickness of 500 $\mu$m for the substrate, and a thickness of 18 $\mu$m for the copper traces. SMA connectors are soldered to the two ends of the microstrip for transmission measurements. To suppress reflection and interference, a gradual transition is made between the corrugated and uncorrugated region.

On our SSPP waveguide, the fingers in the corrugated region has a width of 100 $\mu$m and a length of $3$ mm, with a corrugation period of 500 $\mu$m. The dispersion curve of the fundamental SSPP mode for this design is plotted in Fig.\,\ref{fig1}(b), which is obtained from finite element method (FEM) simulation using COMSOL. The dispersion exhibits a clear deviation from the light line. As the wavevector increases, the frequency of the SSPP mode approaches an asymptotic frequency $f_p$. As a result, both the group and phase velocities are significantly reduced, rendering the SSPP as a slow wave. The asymptotic frequency can be engineering by tailoring the device geometries, and our simulation shows that it is primarily determined by the length of the fingers. For our selection of the finger length ($3$ mm), the asymptotic frequency $f_p=10.5$ GHz, which is favorable for studying the magnon-phonon interaction.

The measured transmission spectrum of the fabricated SSPP waveguide (without the YIG chip) is plotted in Fig.\,\ref{fig1}(c), which  is obtained via S$_{21}$ measurement using a vector network analyzer (VNA). The waveguide shows a transmission band with an insertion loss around 10 dB, which may be attributed to the ohmic loss of copper, the dielectric loss of the substrate, and connector loss. The transmission spectrum shows a cutoff at 10.5 GHz, above which the transmission drops by over 30 dB. As the frequency approaches the cutoff frequency, the phase delay drastically increases because of the reduced phase velocity of the SSPP modes. These measured transmission characteristics agree with the simulated dispersion relation in Fig.\,\ref{fig1}(b). 

The increased phase delay is accompanied by the reduced wavelength of the SSPP, because the wavevector  of the SSPP mode along the propagation direction, $k_x$, increases as the frequency approaches the asymptotic frequency. This also leads to increased value for the evanescent decay constant $k_t=i\sqrt{k_x^2-k_0^2}$ in the transverse directions ($k_0$ is the free-space wavevector), accordingly tighter mode confinement along the transverse direction (e.g., $y$). These are confirmed by the measured SSPP mode profiles, which are plotted in Fig.\,\ref{fig1}(d) for three different frequencies (8.0 GHz, 10.0 GHz, and 10.4 GHz). These mode distributions also agree well with the simulated mode profiles shown in Appendix B.

\begin{figure}[tb]
\includegraphics[width=\linewidth]{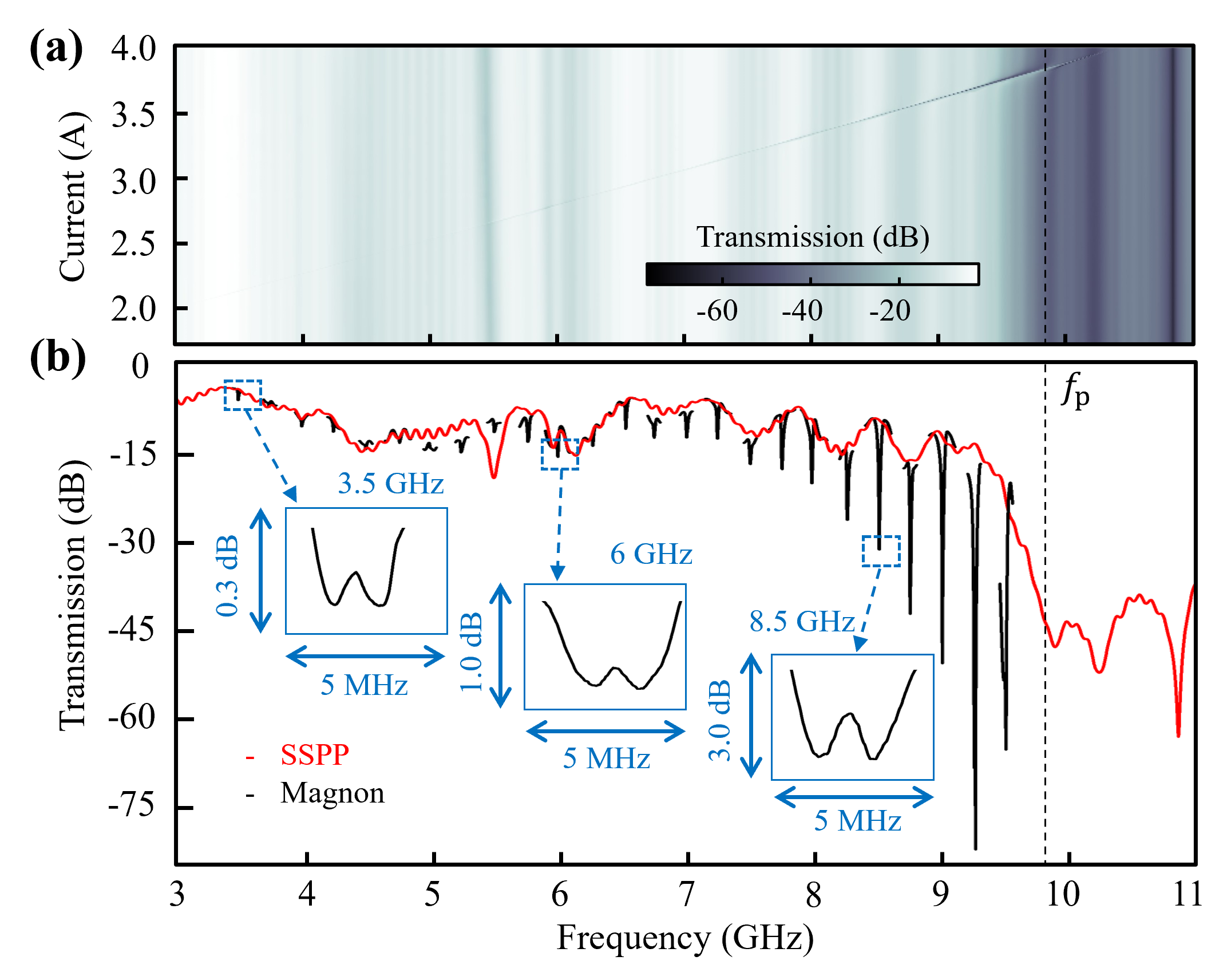}
\caption{(a) Heatmap of the measured SSPP waveguide transmission for Device 1 at different magnet currents. (b) Line plot of the measured SSPP waveguide transmission for the same device, overlaid with magnon resonances at different bias currents. Insets: close-up view of the magnon resonances to show the splitting caused by magnon-phonon coupling.}
\label{fig2}
\end{figure}

On such a device, the microwave signal sent to the device through the SMA connector will first convert to the SSPP mode as it propagates from the uncorrugated microstrip to the SSPP waveguide through the tapered region. When SSPPs propagate through the region covered by the YIG chip, they will interact with the magnon modes in the YIG thin film via magnetic dipole-dipole interaction. Thanks to the small group velocity of the SSPP mode, its interaction with magnons is drastically enhanced compared with guided modes on microstrips, as demonstrated in Ref.\cite{2024_PRL_Xu_slowwave}. In addition, the SSPP modes also exhibit much enhanced mode confinement, further enhancing the magnon-SSPP interaction. As the frequency approaches the asymptotic frequency \( f_p \), the group velocity decreases while mode confinement increases, and consequently, the interaction between SSPPs and magnons gradually intensifies with rising magnon frequency. As a result, larger extinction ratios are observed in the measured SSPP transmission spectra as the bias magnetic field increases, as shown by the transmission spectra for Device 1 in Fig.\,\ref{fig2}(a), where the strength of magnetic field linearly increases with the current applied to the electromagnet. Maximum extinction ratio is observed near $f_p$; When the magnon frequency exceeds $f_p$, magnon mode vanishes from the transmission spectrum because SSPP modes are no longer supported and thus cannot provide enhanced interaction with magnons.

The enhanced SSPP-magnon coupling is more evident in the line plot in Fig.\,\ref{fig2}(b), where the SSPP transmission is superimposed with magnon resonances that are tuned to different frequencies by a series of different bias magnetic fields. The slight deviation in the transmission spectrum as compared with Fig.\,\ref{fig1}(c) is attributed to the presence of the YIG chip which changes the local environment of the SSPP waveguide. At low frequencies, the SSPP mode is very similar to the microstrip mode, with relatively large group velocities and large mode volumes (accordingly, poor mode confinement), which can be inferred from their similar dispersions at low frequencies as shown in Fig.\,\ref{fig1}(b). As a result, the SSPP-magnon coupling is relatively weak, as indicated by the poor extinction ratio of the magnon resonance dips (c.f., at around 4 GHz the extinction ratio is around 1 dB). The extinction ratio increases with frequency, and as the frequency approaches 9.5 GHz (the cutoff frequency for this device), a maximum extinction ratio of nearly 65 dB is achieved, which is attributed to the significantly reduced group velocity and mode volume of the SSPP mode. Importantly, the enhanced coupling of SSPPs with magnons allow the observation of magnon resonances over a broad frequency range (exceeding 7 GHz) with extinction ratios higher than conventional microstrip waveguides. This facilitates the broadband exploration of magnon-phonon interactions, which have typically been limited to a narrow frequency range in conventional cavity magnomechanics due to the finite resonance linewidths of the cavities for accessing the magnon modes.

A zoom-in view of each magnon resonance shown in Fig.\,\ref{fig2}(b) reveals a small splitting on the tip, which results from the coupling of the magnon mode with one HBAR phonon mode. In our device, the whole garnet structure -- 200-nm YIG on top of a 500-$\mu$m GGG substrate -- supports HBAR phonons. Because of the large device thickness, the HBAR modes operating at around 10 GHz have mode orders over 2000. Because of the excellent mechanical properties of YIG and GGG, such HBAR modes have excellent quality factors (typically around 10,000 at room temperature). Note here YIG and GGG have similar mechanical properties, so they will be treated as the same material for phonons hereafter. Thanks to the broadband SSPP-magnon coupling, such splitting induced by the magnon-phonon coupling can also be observed over a 7 GHz frequency range.

\begin{figure}[tb]
\includegraphics[width=\linewidth]{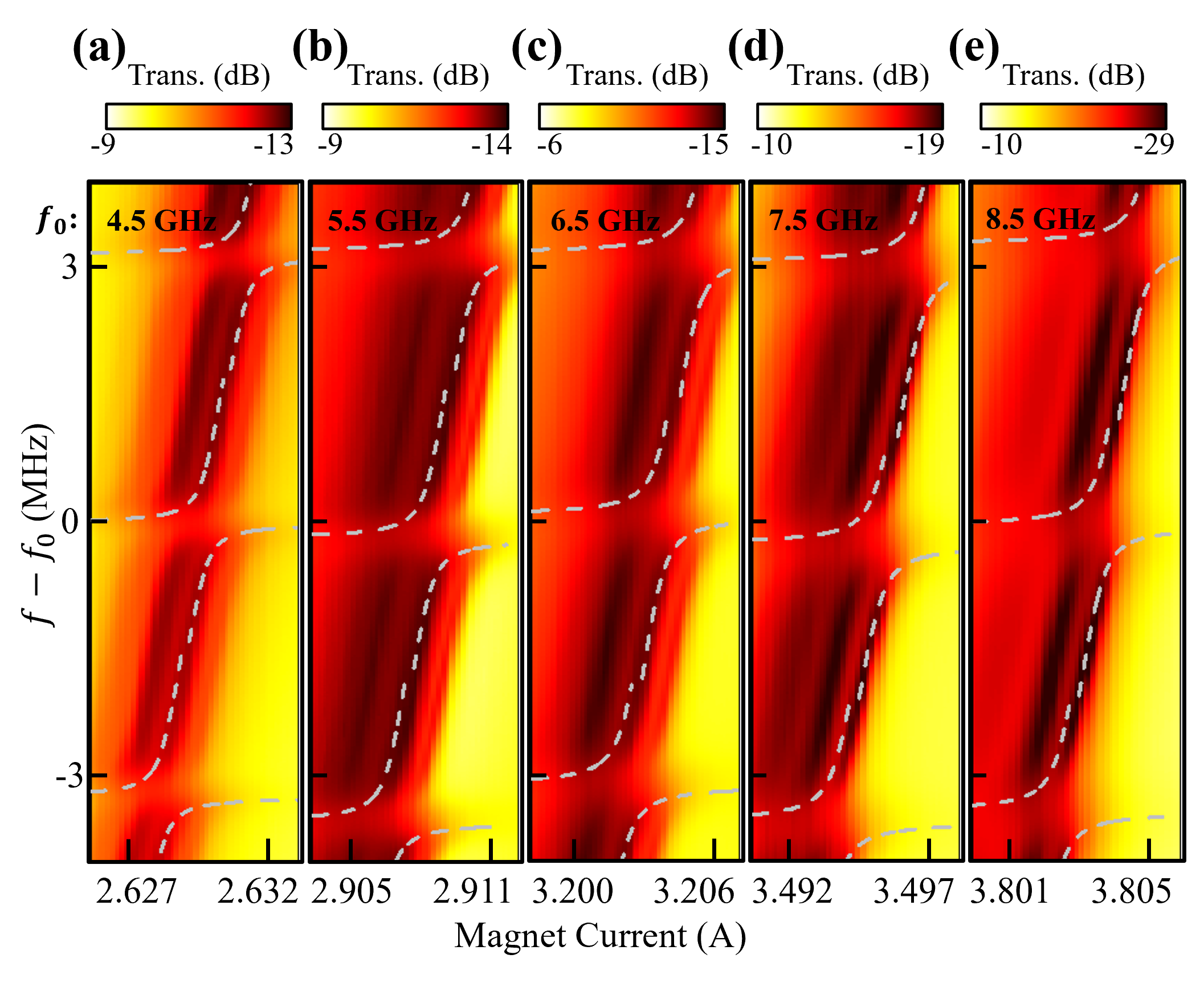}
\caption{Close-up heatmap for measured SSPP transmission, showing the magnon-phonon anti-crossing at selected center frequencies ($f_0$). Trans.: Transmission.}
\label{fig3}
\end{figure}

The three insets in Fig.\,\ref{fig2}(b) show that as the frequency increases, the splitting also becomes larger, indicating stronger magnon-phonon coupling. This is agrees with our initial theoretical prediction, which shows a maximum magnon-phonon coupling can be achieved at 9.8 GHz where best mode volume matching is achieved between the magnon and phonon modes (i.e., the phonon wavelength is twice of the YIG thickness) \cite{2021_PRAppl_Xu_pulseEcho}. When the frequency increases towards 9.8 GHz, better mode overlap can be achieved between the magnon and phonon modes, and therefore it is expected that the magnon-phonon coupling strength increases.

To further verify the acoustic-origin of the splitting, the transmission spectra is measured when the bias magnetic field is swept. The results for Device 2 are plotted in 2-dimensional (2D) heat maps for five selected frequency regimes ranging from 4.5 GHz to 8.5 GHz, as shown in Fig.\,\ref{fig3}. In these measurements, the magnetic field is controlled by the current sent to the electromagent. For each plot, the magnon modes correspond to the tilted lines because their frequency dependes on the bias magnetic field. The horizontal lines correspond to the HBAR phonon modes that are coupled with the magnon modes, which have fixed frequencies without any magnetic field dependence. These horizontal lines are 3.3 MHz apart from one another, matching the calculated free spectral range (FSR) of the HBAR phonons. Whenever a magnon mode becomes on resonance with a phonon mode, avoided crossing is observed. The dashed lines in Fig.\,\ref{fig3} are the calculated frequencies of the coupled modes, which agree well with the measurement results.

\begin{figure}[tb]
\includegraphics[width=\linewidth]{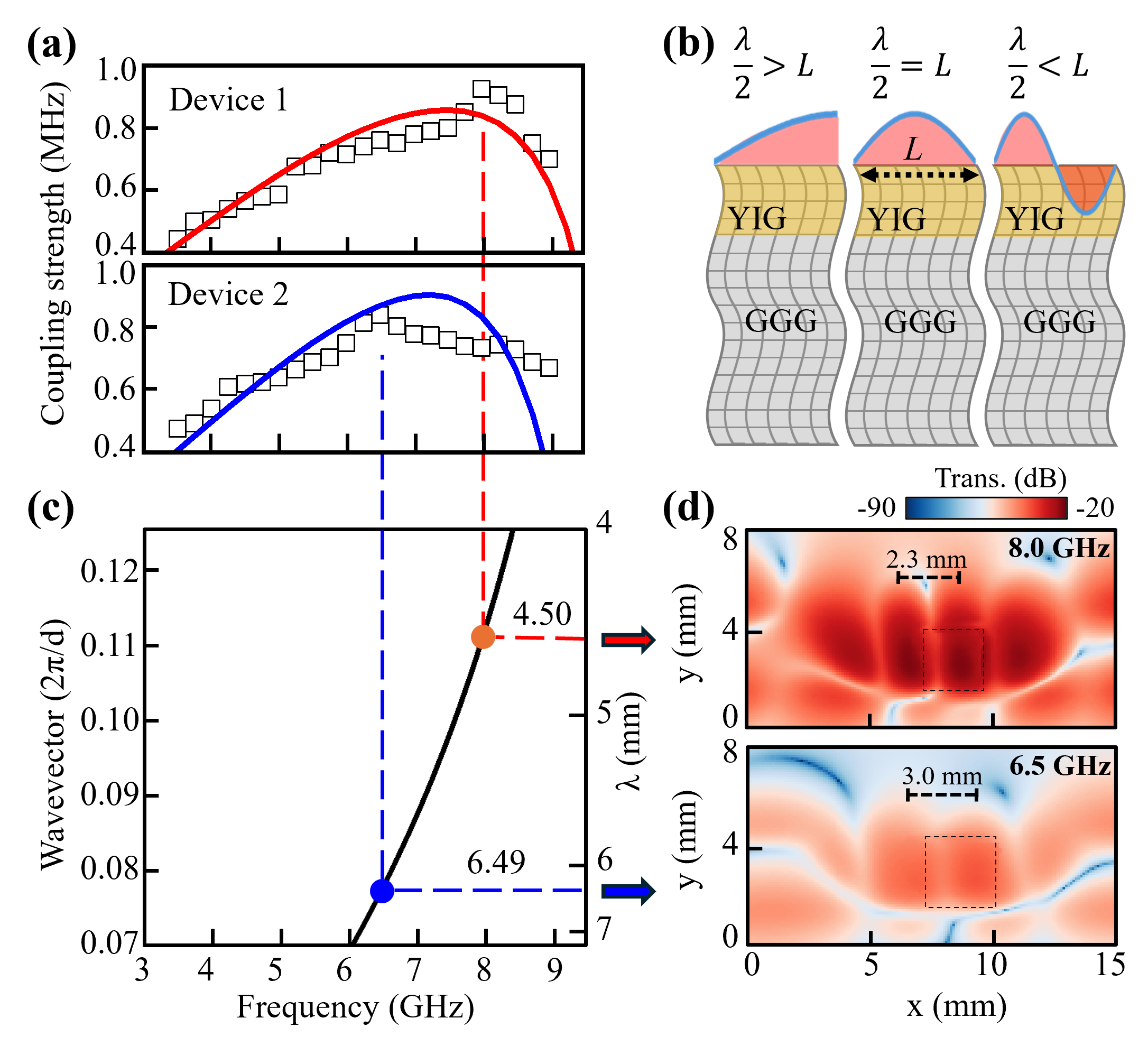}
\caption{(a) Measured (squares) and calculated (solid lines) magnomechanical coupling strength as a function of operation frequency for Device 1 ($2.5\times 2.5$ mm$^2$ YIG film) and Device 2 ($3.0\times 3.0$ mm$^2$ YIG film). (b) Schematic illustration of the size effect. (c) Close-up of the SSPP dispersion, which coverts the peak frequencies in (a) to wavelengths: $\lambda_1=4.50$ mm for Device 1, and $\lambda_2=6.49$ mm for Device 2. (d) Measured mode profiles at the two peak frequencies in (a): 8.0 GHz for Device 1, and 6.5 GHz for Device 2. Dotted squares: position of the YIG thin film device.}
\label{fig4}
\end{figure}

From our calculations, the magnon-phonon coupling strengths at several specific frequencies are derived: 0.62 MHz, 0.69 MHz, 0.84 MHz, 0.76 MHz, and 0.73 MHz for magnon modes at 4.5 GHz, 5.5 GHz, 6.5 GHz, 7.5 GHz, and 8.5 GHz, respectively. Intriguingly, these strengths deviate from the anticipated monotonically increasing trend as frequency rises. To investigate this deviation, mode splitting across the entire 3.5 GHz to 9 GHz frequency band is analyzed for both Device 1 and Device 2. The resulting plot in Fig.\,\ref{fig4}(a) evidently reveals a non-monotonic behavior. As frequency increases, the mode splitting initially rises until it reaches a peak value (8 GHz for Device 1 and 6.5 GHz for Device 2), after which it gradually decreases. This behavior is attributed to the size effect inherent in our YIG devices, as explained below.

In prior studies, a uniform distribution of magnon and phonon modes is often assumed, while their lateral distributions are typically overlooked. Consequently, only the mode overlap along the thickness direction is considered, yielding an optimal coupling frequency of 9.8 GHz for maximum magnon-phonon coupling for 200-nm YIG thin films. This assumption held in earlier demonstrations where large transducers (such as microstrips \cite{An2023Jul} or split ring resonators \cite{2021_PRAppl_Xu_pulseEcho}) efficiently excited the fundamental magnon mode with nearly uniform lateral distributions. However, the highly confined SSPP modes used in this work will excite short-wavelength magnon modes, making it necessary to take into account the size effect of the YIG chips when analyzing the magnon-phonon coupling.

Under the lateral confinement restriction, optimal coupling occurs when the in-plane magnon wavelength satisfies $\lambda_\mathrm{in}/2=L$, where $L$ is the lateral dimension of the YIG chip [Fig.\,\ref{fig4}(b), II]. At lower frequencies [Fig.\,\ref{fig4}(b), I], where the wavelength exceeds the chip dimension, coupling weakens because $\lambda_\mathrm{in}/2>L$. Conversely, at higher frequencies [Fig.\,\ref{fig4}(b), III], wavelength reduction ($\lambda_\mathrm{in}/2<L$) leads to simultaneous existence of positive and negative magnon amplitudes within the finite YIG chip size. Their coupling with the phonon mode will cancel with each other, resulting in reduced magnon-phonon coupling strength. This effect is confirmed by the experimentally mapped mode distributions of the SSPP modes in Fig.\,\ref{fig4}(d), which excites magnon modes with similar profiles. For Device 1, the measurement shows that the SSPP mode at 8.0 GHz has $\lambda_\mathrm{in}/2=2.3$ mm, in agreement with the theoretically predicted wavelength ($\lambda_\mathrm{in}^\mathrm{calc}=4.50$ mm) from the dispersion analysis [Fig.\,\ref{fig4}(c)]. Since it matches the YIG chip size ($L=2.5$ mm) on Device 1, maximum magnon-phonon coupling is achieved at this frequency (8 GHz). Similarly, for Device 2, the SSPP mode at 6.5 GHz has $\lambda_\mathrm{in}/2=3.0$ mm, as shown by the measured mode profile in Fig.\,\ref{fig4}(d), which agrees with the predicted wavelength ($\lambda_\mathrm{in}^\mathrm{calc}=6.49$ mm) and matches the YIG size ($L=3.0$ mm). This corresponds to the optimal mode overlap with the lateral phonon mode, and thus leads to maximum coupling strength at 6.5 GHz for Device 2.

Our experimental observation also matches our analytical calculation based on the model in Appendix A. When the lateral size effect is considered, the calculated coupling strength exhibits a peak for both Device 1 and Device 2, and their peak positions agree with the measurement result [Fig.\,\ref{fig4}(a)]. Although due to nonideality of the experiment and simplification of the model, slight deviations exist between the calculation and measurement results, the non-monotonic dependence of the coupling strength on the operation frequency is unambiguous. On the contrary, if the lateral size effect is ignored, the peak coupling strength will achieved at around 9.8 GHz with a monotonically increasing trend before that, which significantly differs from our experimental observation.

%\section{conclusion}
In conclusion, this work presents a comprehensive investigation of broadband HBAR cavity magnomechanics. By utilizing a SSPP waveguide as the magnon transducer, we observed magnon resonance across a frequency range exceeding 7 GHz, with extinction ratios greater than 1 dB throughout the band. This advancement facilitated the measurement of coherent magnon-phonon interactions throughout this frequency range. The demonstrated broadband magnomechanical interaction significantly surpasses that of conventional devices based on microwave cavities, which typically observe magnon-phonon coupling only within a limited frequency range (usually no more than a few hundred MHz) near the cavity resonance. Thus, our findings pave the way for the development of broadband magnomechanical devices. Furthermore, in comparison to other broadband structures, such as microstrips, our device achieves a significantly enhanced magnon extinction ratio, which is advantageous for applications such as exploring novel nonlinear magnon-phonon interactions. Importantly, the SSPP mode employed in our device enables the excitation of short-wavelength magnons, as opposed to the uniform magnon modes typically utilized in previous magnomechanical devices, resulting in the observation of a novel size effect within our SSPP-magnon-phonon system.Although the device dimensions used in this work are on the millimeter scale, the principle is versatile and can be extended to microscale or nanoscale devices. With the growing interest in developing integrated magnonic circuits, our demonstration offers essential understanding of the fundamental magnon-phonon interaction mechanisms within such hybrid platforms and provide an important engineering toolkit for advancing practical magnomechanical devices. For instance, by precisely tailoring the lateral dimensions of magnonic devices, the size effect can be utilized to suppress magnon-phonon interactions, especially when phonon modes interfere with other interactions involving the magnon mode.

\begin{acknowledgments}
X.Z. acknowledges support from National Science Foundation (2337713) and Office of Naval Research Young Investigator Program (N00014-23-1-2144). 
\end{acknowledgments}

%%%%%%%%%%%%%%%%%%%%%%%%%%%%%%%%%%%%%%%%%%%%%%%%%%%%%%%%%%%%%%%%%%%%%%%%%%%%%%
%%%%%%%%%%%%%%%%%%%%%%%%%%%%%%%%%%%%%%%%%%%%%%%%%%%%%%%%%%%%%%%%%%%%%%%%%%%%%%
%
% Appendices
%
%%%%%%%%%%%%%%%%%%%%%%%%%%%%%%%%%%%%%%%%%%%%%%%%%%%%%%%%%%%%%%%%%%%%%%%%%%%%%%
%%%%%%%%%%%%%%%%%%%%%%%%%%%%%%%%%%%%%%%%%%%%%%%%%%%%%%%%%%%%%%%%%%%%%%%%%%%%%%

\appendix
\section{\textbf{Appendix A: Magnon-phonon coupling strength calculation}}
\label{appendix:a}
\setcounter{equation}{0}
\renewcommand{\theequation}{A\arabic{equation}}

Based on Ref.\,\cite{An2023Jul}, the coupling strength between the magnon mode and phonon mode, with the lateral size effect taken into account, can be calculated from
\begin{equation}
\begin{split}
    \frac{\Omega}{2} \int_0^L\int_{-d}^sm^2dzdx  = \gamma B \int_0^L \int_{-d}^s m\frac{\partial u}{\partial z}dz dx,\\
    \frac{\Omega}{2} \int_0^L \int_{-d}^s u^2 dz dx  = -\frac{B}{2\omega \rho M_s}\int_0^L \int_{-d}^s u\frac{\partial m}{\partial z}dzdx.\\
\end{split}
\label{Eq:NewEq}
\end{equation}
Here $z$ is the out-of-plane direction of the YIG thin film, and the origin ($z=0$) is at the YIG/GGG interface. $m(z,x)=m_1(z)m_2(x)=m_0\theta(-z)\theta(z+d)\cdot \sin[k(x-x_0)+\phi_0]$ is the precessing magnetization of the magnon mode in the YIG film ($-d<z<0$ and $0<x<L$), where $\theta(z)$ is the Heaviside step function, $k$ is the wavevector of the magnon mode,  $x_0$ is the distance of the YIG chip from the end of the SSPP waveguide, and $\phi_0$ is a constant phase accounting for the cable delay. $u=u(z)=u_0\cos{[n\pi(z+d)/(s+d)]}$ is the displacement of the phonon mode throughout the YIG/GGG thickness ($-d<z<s$), where $n$ is the index of the HBAR phonon mode and can be calculated as $n=\frac{2(d+s)}{\lambda}=\frac{2(d+s)\omega}{2\pi v}$, with $v=3850$ m/s being the velocity of shear acoustic waves in YIG. 
$\gamma=28$ GHz/T is the gyromagnetic ratio, $B=7\times 10^5$ J/m$^3$ is the effective magnetoelastic coefficient, $\omega$ is the angular frequency, $\rho = 5100$ kg/m$^3$ is the density of YIG, $M_s=1.374\times 10^6$ A/m is the saturation magnetization of YIG. 

Multiplying these two equations and applying separation of variables lead to:
\begin{equation}
\begin{split}
&\left( \frac{\Omega}{2} \right)^2m_{0}^2{u}_{0}^{2} \frac{d(d+s)}{2} \cdot L \int_0^L m_2^2 dx \\
&=  \frac{-\gamma B^{2}}{2 \omega \rho M_{s}} \int_{-d}^{s} m_1 \frac{\partial u}{\partial z} dz \int_{-d}^{s} u \frac{\partial m_1}{\partial z} dz \left(\int_0^L m_2 dx\right)^2. 
\end{split}
\label{Eq:multiply_2}
\end{equation}

After calculating the integrals, the magnon-phonon coupling strength, with the lateral size effect considered, can be obtained as
\begin{equation}
    \Omega = \Omega_0\cdot F_L,
\end{equation}
where the first term $\Omega_0$ is the coupling strength without considering the lateral size effect, while the second term $F_L$ is the effect of the lateral confinement. 

According to Ref.\,\cite{An2023Jul}, the first term $\Omega_0$ is given as 
\begin{equation}
%\begin{split}
\Omega_0 =2B\sqrt{\frac{\gamma}{\omega\rho M_s d(d+s)}}\left[1-\cos\left(\frac{\omega d}{v}\right)\right].\\    
%\end{split}
\label{Eq:Omega0}
\end{equation}

In our case, the additional term $F_L$ can be calculated as 
\begin{equation}
F_L=\frac{\int_0^Lm_2 dx}{\sqrt{L\int_0^Lm_2^2dx}}=\frac{2\sqrt{2}\sin{\left(\frac{kL}{2}\right)}\sin{\left(\frac{kL}{2}+\phi\right)}}{\sqrt{kL[kL-\sin(kL)\cos{\left(kL+2\phi\right)}]}},
\end{equation}
where $\phi=\phi_0-kx_0$.

Therefore, the coupling strength $\Omega$ with the lateral size effect considered becomes
\begin{equation}
\begin{split}
& \Omega = \Omega_0\cdot \frac{2\sqrt{2}\sin{\left(\frac{kL}{2}\right)}\sin{\left(\frac{kL}{2}+\phi\right)}}{\sqrt{kL[kL-\sin(kL)\cos{\left(kL+2\phi\right)}]}}.
\end{split}
\label{Eq:Omega}
\end{equation} 
The results shown by the solid lines in Fig.\,\ref{fig4}(a) are calculated using this equation. Note that in this calculation, the magnon mode driven by the propagating SSPP mode is treated as propagating wave with the same wavevectors as the SSPP.

\begin{figure}[bt]
\includegraphics[width=\linewidth]{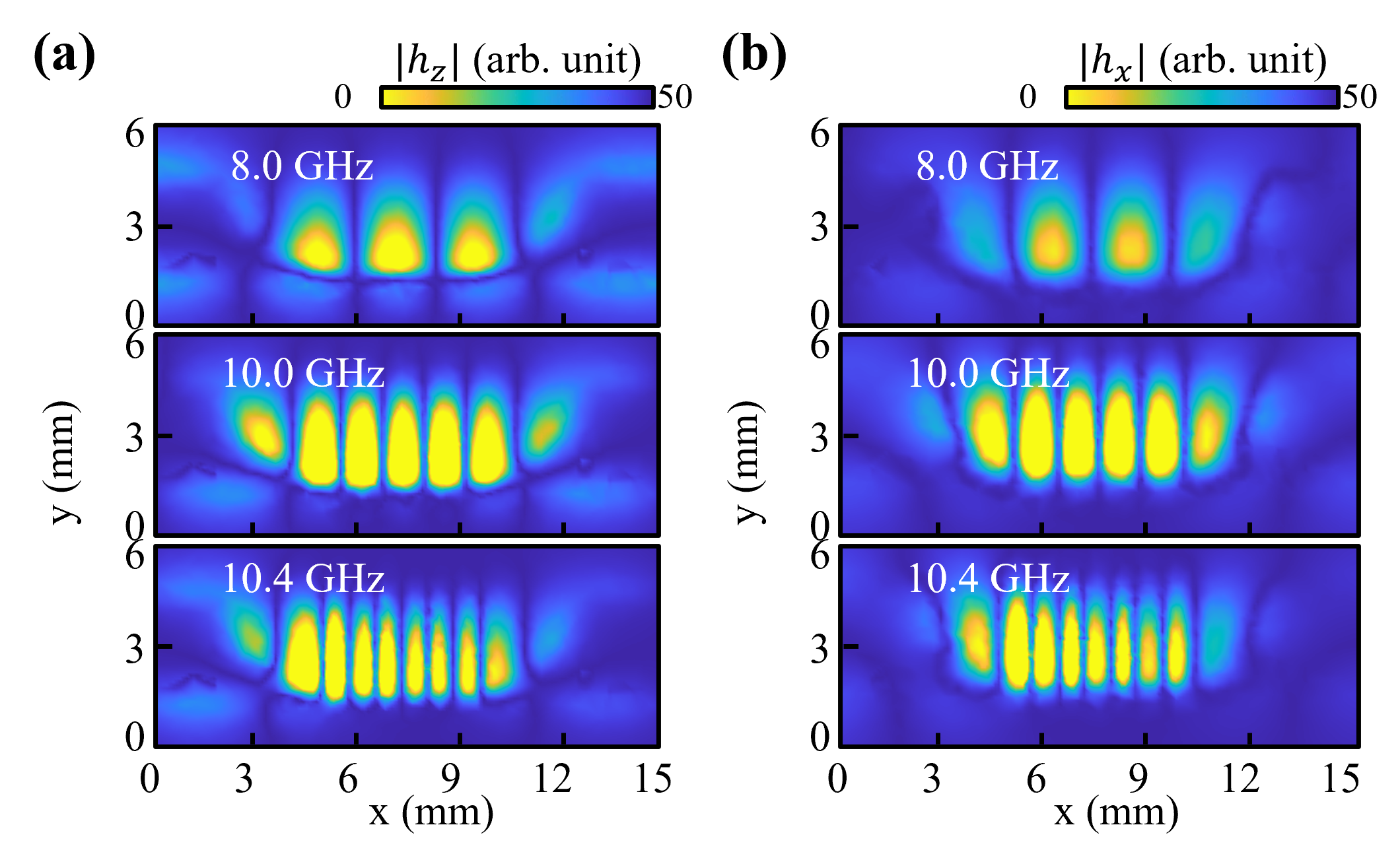}
\caption{Simulated profiles of (a) $h_z$, and (b) $h_x$ of the SSPP mode at three different frequencies, respectively.}
\label{append_fig1}
\end{figure}

\section{Appendix B: Simulated mode profile}
The simulated mode profile of the SSPP mode at three selected frequencies: 8.0 GHz, 10.0 GHz, and 10.4 GHz [corresponding to the plot in Fig.\,\ref{fig1}(d)]. As the frequency increases towards the asymptotic frequency of the SSPP, the wavelength becomes shorter, with more periods showing up. In particular, when the frequency is close to the asymptotic frequency, a slight change in the frequency will lead to a large variation in the wavevector (accordingly the wavelength). From the field plot it can be seen that the $h_z$ and $h_x$ components have the same periods, but their mode profiles are shifted by half a period. In other words, the $h_z$ and $h_x$ components are complementary to each other, with the nodes of the $h_z$ field corresponding to the anti-nodes of the $h_x$ fields. Since the YIG film is magnetized along the out-of-plane direction, only the $h_x$ component interact with the magnon mode.

\bibliography{BBME_arxiv}% Produces the bibliography via BibTeX.

\end{document}